\documentclass[showpacs,amsmath,amssymb,twocolumn,superscriptaddress]{revtex4}
\usepackage{graphicx}

\begin{document}

\title{Size-dependent enhancement of superconductivity in nanowires}
\author{A. A. Shanenko}
\affiliation{Departement Fysica, Universiteit Antwerpen, Groenenborgerlaan 171, B-2020 Antwerpen,
Belgium}
\affiliation{Bogoliubov Laboratory of Theoretical Physics, Joint Institute for Nuclear Research,
141980 Dubna, Russia}
\author{M. D. Croitoru}
\affiliation{Departement Fysica, Universiteit Antwerpen, Groenenborgerlaan 171, B-2020 Antwerpen,
Belgium}
\author{M. Zgirski}
\affiliation{NanoScience Center, Department of Physics, University
of Jyv\"askyl\"a, PB 35, 40014, Jyv\"askyl\"a, Finland}
\author {F. M. Peeters}
\affiliation{Departement Fysica, Universiteit Antwerpen, Groenenborgerlaan 171, B-2020 Antwerpen,
Belgium}
\author{K. Arutyunov}
\affiliation{NanoScience Center, Department of Physics, University of Jyv\"askyl\"a, PB 35, 40014,
Jyv\"askyl\"a, Finland}

\date{\today}

\begin{abstract}
A shape-dependent superconducting resonance can be expected when an energy level associated with the
transverse motion in a wire passes through the Fermi surface. We show that the recently observed
width-dependent increase of $T_c$ in ${\rm Al}$ and ${\rm Sn}$ nanowires is a consequence of this
shape resonance effect.
\end{abstract}


\maketitle
Increasing the critical temperature ($T_c$) of a superconductor (SC) has been a major challenge. On the one
hand one can look for different materials which exhibit a higher $T_c$. Such a search has been very successful
over the last 20 years. On the other hand microstructuring of a superconductor is a different and new road
which is able to modify $T_c$ (i.e. increase and/or decrease) and may also give us further insight in the
basic mechanism of superconductivity.

In earlier works on microstructuring of SC in the mesoscopic regime, enhancement of the critical current
($j_c$) was demonstrated to occur due to trapping of vortices. Also a large increase of the critical
magnetic field ($H_c$) was realized through such mesoscopic structuring which is mostly a consequence of
surface superconductivity. But in both cases the zero-magnetic field critical temperature was unaltered.
The enhancement of $j_c$ and $H_c$ could be accurately described by phenomenological theories such as the
London approach and the (time-dependent) Ginzburg-Landau theory.

In the present Letter we are interested in modifying SC on the nano-scale. We deal with systems where the
electron motion is limited to quasi-one-dimensions (1D). During the last decade nanowires have attracted much
attention in the context of phase fluctuations of the order parameter (i.e. quantum phase slips)~\cite{psl}.
But quantization of the electron motion in the transverse direction was not investigated in much detail.
However, very recently numerical investigation of the Bogoliubov-de Gennes equations has shown~\cite{sc} that
this quantization results in significant shape-dependent superconducting resonances with a profound effect on
the nanowire $T_c$. Such systems have been the subject of recent experimental studies~\cite{tian,fin1,fin2},
and {\it we demonstrate here that the width-dependent increase of $T_c$ found in these experiments is a
manifestation of these shape resonances}.

More than 40 years ago, Blatt and Thompson~\cite{blatt} calculated a remarkable sequence of peaks in the
thickness dependence of the energy-gap parameter of single-crystalline superconducting nanofilms in the clean
limit. They called these spikes shape resonances. At that time it was not possible to produce high-quality
SCs with nanoscale dimensions (only very recently the thickness-dependent oscillations of $T_c$ were observed
experimentally in ultrathin ${\rm Pb}$ films~\cite{guo}). For decades atomic nuclei were the only systems where
the interplay between quantum confinement and pairing of fermions could be studied experimentally and where the
expectations of Blatt and Thompson were confirmed as a series of size resonances in the pairing energy gap of
nuclei~\cite{hilaire}. Very recently high-quality nanowires have become available, where this resonance effect
is expected~\cite{sc} to be significantly enhanced over the two-dimensional $({\rm 2D})$ case.

The physics of the shape-resonance effect in nanowires can be understood as follows. The superconducting order
parameter is not simply the wave function of an ordinary bound state of two fermions but the wave function of a
bound fermion pair in a medium. For example, in the homogeneous situation the Fourier components of the
Cooper-pair wave function are suppressed for wave numbers less than the Fermi wave number due to the presence
of the surrounding Fermi sea~\cite{cooper}. Therefore, the Fourier transform of the Cooper-pair wave function
is essentially nonzero only in the vicinity of the Fermi wave number. In general, for both homogeneous and
inhomogeneous situations, the superconducting order parameter is strongly dependent on $N_D$, the number of
the single-electron states (for one spin projection) situated in the Debye "window" around the Fermi level $[\mu-
\hbar \omega_D, \mu +\hbar \omega_D]$, where $\mu$ is the chemical potential (the Fermi energy) and $\omega_D$
is the Debye frequency~\cite{fetter,degen}. More precisely, the mean energy density of these states per volume
unit $N_D/(2\hbar\omega_D V)$ is of importance. When a SC is small enough, then this density varies together with
its characteristic size. In particular, in the presence of quantum confinement the band of single-electron states
in a clean single-crystalline wire is divided up in a series of parabolic one-dimensional subbands. While
the width (the cross section $\sigma$) of a wire increases/decreases, these subbands move down/up in energy.
Each time when the bottom of a subband enters the Debye window, the density $N_D/(2\hbar\omega_D V)$ increases.
This results in a sequence of peaks in the density of states versus $\sigma$, i.e., the shape resonances. The
resonances are significant for nanowires but are smoothed out with increasing nanowire width. In the large-$
\sigma$ limit the density $N_D/(2\hbar\omega_D V)$ is nearly constant and slowly approaching the well-known bulk
value $N(0)= mk_F/(2\pi^2\hbar^2)$~(see, for example, Ref.~\cite{fetter}), where $k_F$ is the $3D$ Fermi wave
number. It is of importance that the shape resonances in the density of states are accompanied by significant
peaks in the $\sigma-$dependence of the averaged (over spatial coordinates) order parameter~\cite{sc}. This
average is close to the energy gap in the quasiparticle spectrum, whereas the latter (taken at zero temperature)
is proportional to $T_c$. Hence, the critical SC temperature in a nanowire is an oscillating function of $\sigma$
in the presence of shape resonances (these oscillations will be smoothened by fluctuations in the cross section).

To explore shape resonances, one needs to deal with a single-crystalline SC (or with a polycrystalline SC made of
strongly coupled grains) where the electron mean free path is comparable with the SC characteristic size. In the
considered nanowires this path is at least of the order of the wire width~\cite{fin1,fin2}. Such a nanowire is
in the clean limit in the transverse direction: impurities influence only the electron motion in the longitudinal
direction. Use of this approximation together with arguments similar to the Anderson theorem~\cite{degen} makes
it possible to conclude that nonmagnetic impurities do not produce a significant effect on the order parameter
in this regime. The Bogoliubov-de Gennes (BdG) equations~\cite{degen} are the microscopic equations describing a
clean SC when the order parameter $\Delta({\bf r})$ changes with position. In the absence of a magnetic field,
$\Delta({\bf r})$ can be chosen as a real quantity (phase effects are beyond the scope of the present work),
and the BdG equations become
\begin{eqnarray}
E_i u_i({\bf r})= \Bigl(-\frac{\hbar^2}{2m} \nabla^2 -\mu\Bigr)u_i({\bf r}) + \Delta({\bf r})
                                                                                     v_i({\bf r}),
\label{BdG1}\\ [1mm]
E_i v_i({\bf r})= \Delta({\bf r}) u_i({\bf r})- \Bigl(-\frac{\hbar^2}{2m} \nabla^2 -\mu\Bigr)
                                                                                     v_i({\bf r}),
\label{BdG2}
\end{eqnarray}
where $E_i$ is the quasiparticle spectrum, $\mu$ stands for the chemical potential and $m$ denotes the
electron band mass. Equations (\ref{BdG1}) and (\ref{BdG2}) are supplemented by the self-consistency
relation $\Delta = g\,\sum_i\,u_i v_i^*(1-2f_i)$ with $g$ the coupling constant and $f_i=f(E_i)$ the
Fermi function. The summation is over all eigenstates with the kinetic energy within the Debye window
$[\mu-\hbar\omega_D,\mu+\hbar \omega_D]$. This kinetic energy is very close to the single-electron
energy in a gas of unpaired electrons. The chemical potential is determined by $n = (2/V)\sum_i\int
d^3r[|u_i|^2 f_i+|v_i|^2(1-f_i)]$ with $n$ the mean electron density. Due to the quantum confinement
in the transverse directions we have to set $u_i({\bf r})|_{{\bf r}\in S} = v_i({\bf r})|_{{\bf r} \in
S}=0$ on the wire surface while in the longitudinal direction periodic boundary conditions are used.

\begin{figure}[t]
\centerline{\includegraphics[width=6.2cm]{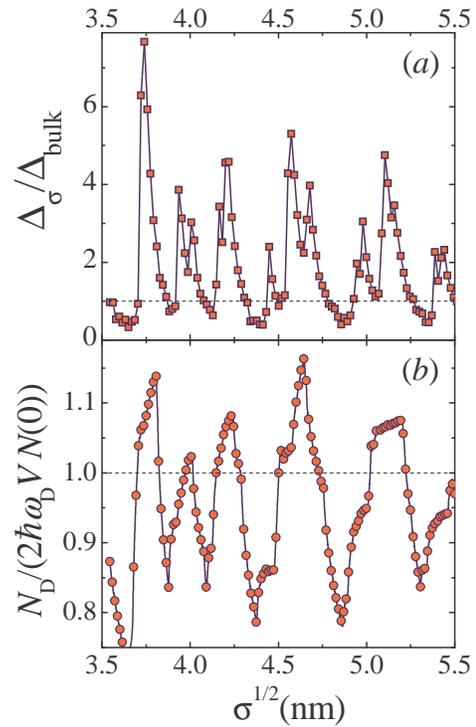}}
\vspace{-0.2cm}
\caption{(Color online) The width-dependent relative gap $\Delta_{\sigma}/\Delta_{{\rm bulk}}$ (a)
and the relative mean density of the single-electron states in the Debye window $N_D/(2\hbar \omega_D
V N(0))$ (b) versus $\sigma^{1/2}$ for cylindrical ${\rm Al}$ nanowires with cross section $\sigma$
at zero temperature.}
\vspace{-0.5cm}
\label{fig1}
\end{figure}

Fig.~\ref{fig1} shows the numerical results of Eqs.~(\ref{BdG1}) and (\ref{BdG2}) for a cylindrical wire with length
$L=1 {\rm \mu m}$ for ${\rm Al}$ parameters: $gN(0)=0.18$ and $\hbar\omega_D /k_B = 375\,{\rm K}$~\cite{fetter,degen}.
The bulk chemical potential is chosen as $\mu_{\rm bulk}=0.9\;{\rm eV}$~(it corresponds to $n=3.88\cdot 10^{21}
{\rm cm}^{-3}$, see the remark about influence of the electron-density choice below) and the electron band mass $m$
is taken to be the free electron mass. In Fig.~\ref{fig1} (a) the zero-temperature energy gap $\Delta_{\sigma}$
normalized by its bulk value $\Delta_{\rm bulk}$ is plotted, whereas in panel (b) the mean density of the
single-electron states in the Debye window $N_D/(2\hbar\omega_D V)$ is shown versus the effective nanowire diameter
$\sigma^{1/2}$ in units of $N(0)$. Notice that shape resonances in $N_D/(2\hbar\omega_D V N(0))$ are accompanied by
large oscillations in $\Delta_{\sigma}/\Delta_{\rm bulk}$. Amplitudes of these oscillations are an order of magnitude
larger as compared to similar peaks in the gap function of thin films with the same width~\cite{blatt,chen}. We can
point out, at least, three reasons for such enhancement. First, quantum confinement is much stronger in nanowires.
Second, contrary to a film, most single-electron subbands of a nanowire are degenerate. For example, in the situation
of a cylindrical nanowire there are three quantum numbers for $u_i$ and $v_i$: $i=\{j, m_{\varphi}, k\}$ with $j$
the radial quantum number, $m_{\varphi}$ the azimuthal quantum number, and $k$ the wave number of the free longitudinal
motion. Any energy subband with $m_{\varphi} \not=0$ is degenerate due to $E_{j,m_{\varphi},k}= E_{j, -m_{\varphi},k}$.
Third, even if the subbands are not degenerate, they can be situated very close to one another in energy (in particular,
this is typical for subbands with large quantum numbers). In this case a resonance has not decayed when the next one
appears, and they join in one profound peak. Note that in thin films the superconducting shape resonances are well
separated from each other and equidistant~\cite{blatt,chen} which is clearly not the case here. Another important point
about Fig.~\ref{fig1} is that there is an important difference between dependencies presented in panels (a) and (b) of
Fig.~\ref{fig1}. The energy gap is strongly enhanced at the resonant points (with respect to the bulk value) while there
are much less important drops between the resonances. This is different from the density of states where the oscillations
are almost centered around the bulk value. The reason is that the resonant increase of $N_D/(2\hbar\omega_D V)$ occurs due
to states making practically the same contribution to the order parameter, and this collective mechanism plays the role
of an "amplifier" (this is why the word "resonance" is appropriate here). Indeed, we have $u_i\propto e^{ikz}$ and $v_i
\propto e^{ikz}$, where $z$ is the longitudinal coordinate. Hence, the product $u_i v^*_i$ appearing in the
self-consistency relation for the order parameter does not depend on $k$ and remains the same for a given subband.
But this product strongly oscillates when the quantum numbers describing the transverse states are changed
which destroys the collective behavior beyond a resonant point.

\begin{figure} [t]
\centerline{\includegraphics[width=7.4cm]{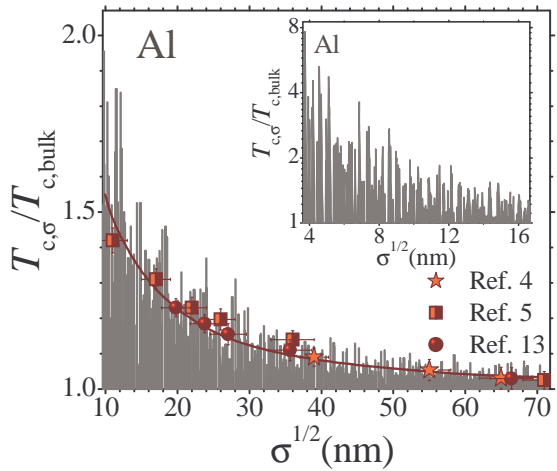}}
\vspace{-0.3cm}
\caption{(Color online) Critical temperature $T_{c,\sigma}/T_{c,{\rm bulk}}$ versus the square root of the
cross section of a cylindrical ${\rm Al}$ nanowire. Symbols are the experimental results from Refs.
\cite{fin1,fin2,fin3}, the solid curve is a guide to the eye and indicates the general trend of the
experimental results. The inset: the same dependence but in the small-$\sigma$ region.}
\vspace{-0.2cm}
\label{fig2}
\end{figure}

\begin{figure}[t]
\centerline{\includegraphics[width=6.4cm]{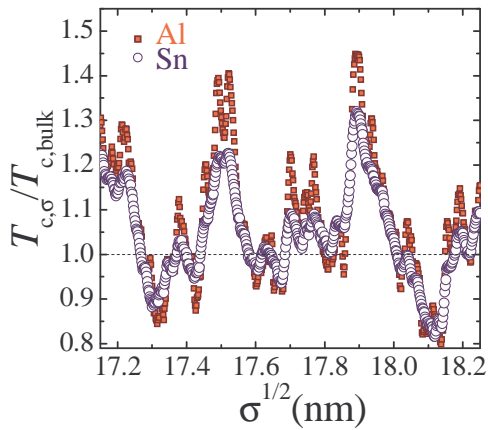}}
\vspace{-0.4cm}
\caption{(Color online) The shape resonances in the relative transition temperature for ${\rm Al}$ and
${\rm Sn}$ cylindrical nanowires.}
\label{fig3}
\end{figure}

The resonant peaks in the energy-gap function are much higher than the bulk value, whereas drops of $\Delta_{
\sigma}$ between resonances are not so significant. The same is true for the critical temperature. This is the
reason why the experimentally measured averaged critical temperature increases monotonically with decreasing
wire width in the presence of inevitable cross section variations in real samples.

Fig.~\ref{fig2} shows the effect of the shape resonances on the transition temperature $T_{c,\sigma}/T_{c,{\rm
bulk}}$ for cylindrical ${\rm Al}$ nanowires. We calculated $T_c$ for discrete values of $\sigma^{1/2}$ with
steps of $0.01{\rm nm}$. The nanowire length and the mean electron density are the same as in Fig.~\ref{fig1}.
The theoretical results are compared with recent experimental data for ${\rm Al}$ polycrystalline nanowires
with strongly coupled grains. The three sets of experimental results correspond to three different initial
samples having the transverse dimensions progressively reduced by the ion-beam sputtering method~\cite{fin1}.
The squares~\cite{fin2} and circles~\cite{fin3} represent the observations made for two samples fabricated on
the same chip, whereas the data given with the stars~\cite{fin1} are from another experiment. The horizontal
bars are due to a distribution in the wire cross section. The vertical error bars are a consequence of
uncertainties in the choice of the bulk superconducting temperature due to the presence of impurities: the
upper limit corresponds to the transition temperature in clean aluminum $T_{c,{\rm bulk}}= 1.19\,{\rm K}$,
the lower limit $T_{c,{\rm bulk}}=1.26\,{\rm K}$ is used for the squares and circles (this is the transition
temperature in the nanowire with $\sigma^{1/2}= 71\;{\rm nm}$, the widest in these two sets) and $T_{c,{\rm
bulk}} = 1.27\,{\rm K}$ for the stars ($T_{c,\sigma}$ for the nanowire with $\sigma^{1/2}=65\;{\rm nm}$, the
widest in the set of Ref.~\cite{fin1}). The experimental transition temperatures shown in Fig.~\ref{fig2} were
extracted from the temperature dependent resistance as the point at which the resistance dropped to one half of
its normal-state value. From Fig.~\ref{fig2} it follows that the general trend of the experimental results is in
good agreement with the calculated resonance amplitudes and is close to the average trend of resonances.

The same conclusion holds for nearly single-crystalline ${\rm Sn}$ nanowires fabricated with the method of
electrodeposition into porous membranes~\cite{tian}. Though there is no detailed information available about
$T_{c,\sigma}$ in this case, we have at our disposal a clear signature of its increase up to $1.1\;T_{c,{\rm
bulk}}$ in a cylindrical tin nanowire with a width of about $20\;{\rm nm}\;(\sigma^{1/2} \approx 17.7\;{\rm
nm})$~\cite{tian}. Using the experimental trend from Fig.~\ref{fig2} one obtains $T_{c,\sigma}/ T_{c,{\rm bulk}}
\approx 1.25$ for an aluminum nanowire with the same width. Hence, the size-dependent effect on the critical
temperature is a factor of $2$ smaller when passing from ${\rm Al}$ to ${\rm Sn}$ for a nanowire with an effective
diameter of about $\sigma^{1/2} \approx 17.7 {\rm nm}$. In Fig.~\ref{fig3} the transition temperature calculated
with the BdG equations for a cylindrical nanowire is shown for ${\rm Al}$ and ${\rm Sn}$~($gN(0)=0.25$ and $\hbar
\omega_D /k_B = 195\,{\rm K}$~\cite{fetter,degen}) near $\sigma^{1/2}=17.7 \;{\rm nm}$. One can see that the
amplitudes for ${\rm Sn}$ are indeed smaller, and the reduction factor changes between $1.4$ and $2.5$ (depending
on the specific resonance).

These findings show that the shape-reso\-nance effect plays an essential role in the size-dependent increase of
the critical SC temperature recently observed in experiments with ${\rm Al}$ and ${\rm Sn}$ nanowires. Note that
the resonant amplitudes are weakly dependent on the total electron density in the metallic domain $n \approx
10^{21}-10^{23}\;{\rm cm}^{-3}$. For example, when the density rises from $3.88 \cdot10^{21}\;{\rm cm}^{-3}$ to $2
\cdot 10^{22} \;{\rm cm}^{-3}$, the amplitudes of the most profound shape resonances are reduced by $10\%- 15\%$ while
the change in the "ordinary" resonant deviations from the bulk value is practically negligible~\cite{sc}. The mean
distance between two neighboring resonances is determined by the inverse 3D Fermi wave number and is therefore very
sensitive to the electron density $n$. For instance, instead of $1-2$ resonances per $0.2\; {\rm nm}$ for $n = 3.88
\cdot 10^{21}\;{\rm cm}^{-3}$, we get $3-4$ resonances for $n = 2 \cdot 10^{22}\;{\rm cm}^{-3}$~\cite{sc}. Thus, the
concrete value for $n$ in the metallic domain is not very important for the main result of the present work.

Concluding, we have shown that the size-dependent increase of the superconducting temperature recently found in
${\rm Al}$ and ${\rm Sn}$ nanowires is well explained by the shape-resonance effect. The surface-phonon mechanism
\cite{naugle}, which is usually considered to be responsible for the $T_c$-enhancement in granular films and wires
(note that this enhancement is insensitive to the cross section!)~\cite{gran}, can also contribute but in our
systems it is of secondary importance. For example, for ${\rm Al}$ and ${\rm Sn}$ films with thickness of about
$5\,{\rm nm}$ this contribution can be estimated to be $\approx 0.1\,{\rm K}$~\cite{naugle}. While the amplitudes
of the shape resonances are about $1 {\rm K}$ for this film thickness~(see Refs.~\cite{blatt} and~\cite{chen}).

This work was supported by the Flemish Science Foundation (FWO-Vl), the Belgian Science Policy, BOF-TOP
(University of Antwerp), and by the EU Commission FP6 NMP-3 Project No. 505457-1 ULTRA-1D.


\end{document}